\documentclass[runningheads]{llncs}
\usepackage{subcaption}
\usepackage{graphicx}
\usepackage{comment}
\usepackage{amsmath,amssymb} 
\usepackage{color}

\definecolor{forestgreen}{rgb}{0.15, 0.55, 0.125}

\newcommand{\reals}{\mathbb{R}}

\usepackage[colorinlistoftodos]{todonotes}
\UseRawInputEncoding

\begin{document}
\title{3D Convolutional Sequence to Sequence Model for Vertebral Compression Fractures Identification in CT}
\titlerunning{3D Convolutional Sequence to Sequence Model}
%
%
\newcommand*\samethanks[1][\value{footnote}]{\footnotemark[#1]}
\author{David Chettrit\inst{1}\thanks{Equal contribution} \email{davidvcfpaper@outlook.com} \and
Tomer Meir\inst{2}\samethanks \and
Hila Lebel\inst{3} \and
Mila Orlovsky\inst{3} \and
Ronen Gordon\inst{3} \and
Ayelet Akselrod-Ballin \inst{3}\thanks{Equal advising}  \and 
Amir Bar\inst{3}\samethanks[2]}


%
\authorrunning{ D. Chettrit et al.}

%
\institute{ Independent AI Consultant \and Weizmann Institute of Science \and Zebra Medical Vision Ltd}

\maketitle              

\begin{abstract}
An osteoporosis-related fracture occurs every three seconds worldwide, affecting one in three women and one in five men aged over 50. The early detection of at-risk patients facilitates effective and well-evidenced preventative interventions, reducing the incidence of major osteoporotic fractures. 
In this study we present an automatic system for identification of vertebral compression fractures on Computed Tomography images, which are often an undiagnosed precursor to major osteoporosis-related fractures. The system integrates a compact 3D representation of the spine, utilizing a Convolutional Neural Network (CNN) for spinal cord detection and a novel end-to-end sequence to sequence 3D architecture. We evaluate several model variants that exploit different representation and classification approaches, and present a framework combining an ensemble of models that achieves state of the art results, validated on a large data set, with a patient-level fracture identification of 0.955 Area Under the Curve (AUC). The system proposed has the potential to support osteoporosis clinical management, improve treatment pathways and to change the course of one of the most burdensome diseases of our generation. 

\end{abstract}

\section{Introduction}
Worldwide, an osteoporotic fracture occurs on average every three seconds, with one in three women and one in five men aged over 50 experiencing an osteoporotic fracture \cite{Johnell_2006,Kanis_2000} at significant economic cost to health and social care systems. Treatments of osteoporosis are widely available, evidence-based and cost effective \cite{Kanis_2005}, but a considerable diagnostic gap exists to identify patients at risk of fracture. However, in over 55\% of cases, major osteoporotic fractures (MOFs) are preceded by an often asymptomatic and undiagnosed warning sign, the vertebral compression fracture (VCF) \cite{Hasserius_2003,Edwards_2007}.
\par Despite their prognostic value, VCFs commonly go unreported on routine radiological imaging, with between 13-16\% of retrospectively confirmed VCFs actually reported at the time of the initial imaging study \cite{carberry_2013,williams_2009}. Typically, VCFs are not a difficult radiological diagnosis, rather they are overlooked as ‘incidental’ findings relative to the primary reason for undertaking the CT examination \cite{bartalena_2010}, and some automatic approaches have been proposed to mitigate this \cite{krishnaraj2019simulating,dagan2020automated}. 

Previous work in the domain of automatic identification of VCF utilize traditional machine learning approaches, which commonly focus on vertebra segmentation and are applied on small data sets of several hundred of studies (\cite{jakubicek_2020},\cite{Ramos_2019}). For example, the work by Valentinitsch et al. \cite{valentinitsch_2019}  employed a random forest (RF) classifier based on texture features and regional vertebral Bone Mineral Density analysis to identify VCF, and was applied on approximately 200 patients. 

\par Recently, deep learning has been used for the automatic detection of various findings in medical images \cite{hamidian20173d,laserson2018textray,chettrit2019pht,bar2019improved}. For the automatic detection of Compression Fractures, Chen et al. \cite{chen_2015} proposed a Convolutional Neural Network (CNN) model with an initial vertebra localization using RF's. The authors in Bar et al. \cite{bar_2017}, first segment the portion of the spine included in a Computed Tomography (CT) scan and virtually construct a 2D sagittal section of the spine. Then a two-dimensional (2D) CNN classifies the sagittal patches and finally a Recurrent Neural Network (RNN) is applied on the resulting vector of probabilities. Similarly, Tomita et al. \cite{tomita_2018} proposed a 2D approach that first utilizes a CNN-based feature extraction module from 2D CT slices and then uses an RNN based feature aggregation method based on Long Short-Term Memory (LSTM) networks. 
RNN's and specifically LSTM's have been successfully applied in many applications \cite{sherstinsky_2020}. Still, this type of representation is limited in exploiting the full vertebra sequence information composing the spine. 
\par A different approach by Sekuboyina et al \cite{sekuboyina_2018} performed vertebra labeling based on a butterfly-shaped network architecture that combines the information from the sagittal and coronal maximal intensity projection images in addition to an adversarial energy-based auto-encoder as a discriminator for learning of the butterfly net.  Our proposed system differs from the majority of studies in the field that rely on 2D/2.5D in that it uses 3D vertebral volumes.
Similarly, a recent study by Nicholas et al. \cite{nicolaes_2020} presented a two staged VCF detection method that first predicts the class probability for every voxel using a 3D CNN, followed by aggregation of the information to VCF and patient level predictions. This study demonstrated the advantage of 3D representation, yet its major shortcoming was the use of a small dataset, of only 90 patients.

To summarize, this study proposes an automatic system for identification of VCF which integrates several novel deep learning components. Our contribution is three fold. 
First, we create a compact 3D representation of the spine. The representation is based on a spinal cord detection CNN, a reconstructed sagittal view, and volumetric patches along the vertebral columnn. The benefit of this representation is that it preserves 3D details, important for diagnosis, incorporating intensity, texture and localization features, making it effective for further classification. Second, to the best of our knowledge we are the first to combine a sequence to sequence architecture learned end-to-end for this type of application. Leveraging a sequence of 3D features based on a sequence of vertebrae allows incorporating local 3D features with the entire vertebra column information. Third, we report state of the art (SOTA) results, on one of the largest CT datasets in this problem domain. Finally, the system presented here can be used in clinical practice for early detection of patients suspicious for osteoporosis, enabling focused interventions and reducing the incidence, thus fracture burden on a wider scale.

\section{Methods}
\begin{figure}[t]
\centering
\includegraphics[scale=1.3]{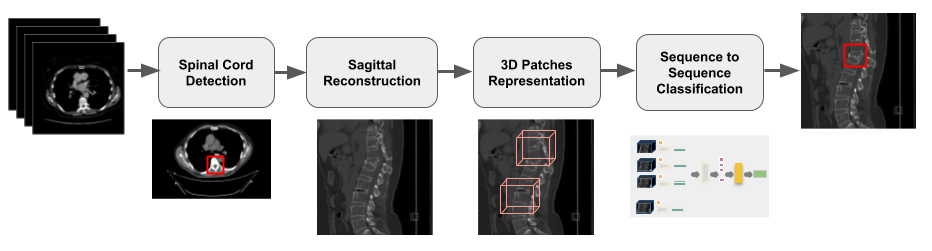}
\caption{System outline. Given a CT scan, the algorithm combines 3D components to output a reconstructed sagittal view  detecting VCF in a bounding box.}

\label{fig:sysOutline}
\end{figure}

The framework inputs chest and abdominal axial CT images and generates a compact 3D representation of the spine. This phase includes spinal cord detection based on a CNN, sagittal reconstruction, and extraction of 3D patches along the vertebral column.  Then, a sequence to sequence compression fractures classification and localization component is applied. A high level description of our method is depicted in Figure~\ref{fig:sysOutline}.

\subsection{From CT to Vertebral Column Region of Interest (ROI)}
A YOLO \cite{redmon_2017} detector type of CNN was utilized in order to localize the spinal cord on axial CT scans, allowing an accurate definition of the region of interest (ROI) for the pipeline. The detector was trained to identify the spinal cord location on every axial slice in the volume each 30 mm. The remaining locations were then linearly interpolated. The model was trained on 300 slices from the training set where the spinal cord location was annotated by a bounding box. 

\subsection{From Vertebral Column ROI to Patch Representation}
\label{section:roi_to_patch}

Due to the large data size involved in CT imaging of the spine, a 3D compact representation of the spine VOI (volume of interest) is created as follows.  
First, the sagittal view is reconstructed from the axial scans of the volume. Each sagittal slice in the 3D volume is resampled to maintain equal pixel size in x and y. The resampling is done along the column of each slice column using Fourier Transforms. Then, given the spinal cord localization, the volume surrounding the spinal column is cropped and partitioned into a set of $k$ 3D volumetric patches, such that the vertebra column is completely tiled with a minimal overlap. The patches are sampled, resized to $(p_{H},p_{W},p_{Z})$ corresponding to the size of the patches in the sagittal, coronal, and axial dimensions respectively, and normalized after a HU windowing is applied (window center 370 HU and window width 840 HU). This step produces the input $(k,p_{H},p_{W},p_{Z},1)$ to the following classification and localization step.

\begin{figure}[t]
\centering
\includegraphics[scale=1]{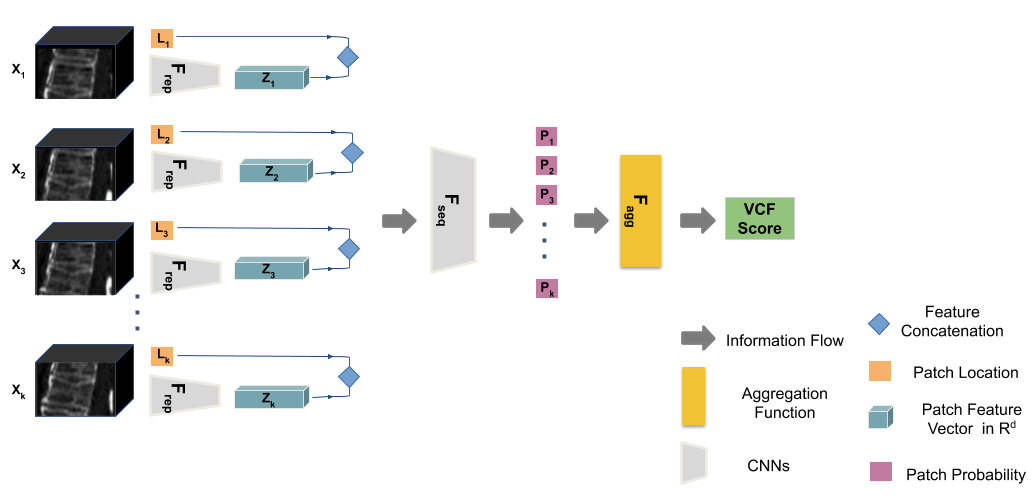}
\caption{Flowchart of the classification and localization architecture. It presents the flow from 3D input patches ($X_{i}$) to the series overall score (VCF score), including the patch feature extractor $F_{rep}$, the position information per patch ($L_{i}$), the sequence to sequence CNN $F_{seq}$, the fracture probabilities per patch ($P_{i}$), and the aggregation function $F_{agg}$.}
\label{fig:Seq2Seq}
\end{figure}

\subsection{Classification and Localization}
A high level description of this classification and fracture localization method is depicted in Figure~\ref{fig:Seq2Seq}.\\
\par \textbf{Basic Setup}. Let $X=(x_1, ... , x_k)$ be an input sequence of length $k$, where every $x_i \in \reals^{H \times W \times Z}$ is a single $3D$ patch of the representation described in \ref{section:roi_to_patch}. Next, we use three function: $F_{\text{rep}}$, $F_{\text{agg}}$ and $F_{\text{seq}}$, where $F_{\text{rep}}$ is a mapping from $\reals^{H \times W \times Z}$ to $\reals^{D}$ and is typically used to map a patch into a new vector representation; $F_{\text{agg}}$ is a mapping from $\reals^{K \times D}$ to $\{0,1\}$, e.g. performs aggregation of an entire sequence into a single result; and $F_{\text{seq}}$ is a mapping from $\reals^{K \times D}$ to $\{0,1\}^{K}$, e.g, performs classification for each sequence item, considering other sequence items. $F_{\text{rep}}$, $F_{\text{agg}}$ and $F_{\text{seq}}$ are typically learned neural networks. Thus, given an input sequence $X$ we start by applying $F_{\text{rep}}$ to obtain an input representation per sequence item:
$${Z} = (F_{\text{rep}}(x_1), ..., F_{\text{rep}}(x_k))$$

Where $Z_i \in \reals^{D}$ is the representation obtained for each sequence item. We then use $F_{\text{agg}}$ and $F_{\text{seq}}$ to obtain an aggregated sequence result and a result per sequence item respectively. To do this, we simply need to apply $F_{\text{agg}}$ and $F_{\text{seq}}$.
Our loss is thus defined as follows:

$$L = L_1(Y_{agg}, F_{\text{agg}}(F_{\text{seq}}(Z))) + \lambda L_{2}(Y_{seq}, F_{\text{seq}}(Z))$$

Where we set $L_1$ and $L_2$ as the binary cross entropy loss. $Y_{agg}$ and $Y_{seq}$ are the ground truth fracture label per series and per patch, respectively. $\lambda$ is here a weight to the sequence component of the loss. We train this system in an end-to-end fashion. Next, we explore possible choices for $F_{\text{rep}}$, $F_{\text{seq}}$ and $F_{\text{agg}}$.

\textbf{Patch Representation.} As described, the input patch representation is typically a cropped $3D$ region of a CT image. Thus, we can obtain a natural representation of it using a $3D$ CNN. Specifically, we use a CNN of the following architecture: 3 blocks, where each blocks contains two 3D convolutions followed by 3D Max Pooling. The convolution filters is doubled for every block, and is initially starting at 8. The resulting output is used a feature vector representation per patch.

We note that a single patch is independent of the global context of the entire volume and of the specific location where it originally resides. To alleviate this difficulty, we supplement the learned CNN representation with a location descriptor. This enables learning a representation obtained by the CNN. A location feature is simply a scalar indicating the anatomical relative position of the patch center.

\textbf{Sequence Items Classification.} Thus far we have considered ways to map an input sequence of patches into a new sequence of representation which better captures the higher level semantics in the input sequence. Next, we describe possible ways to binary classify single sequence items. Perhaps the most natural strategy would be to convolve a $max$ filter over the sequence such that every sequence item is scored according to its environment $max$ value. An alternative could be to go over the sequence in a serial manner, e.g. top to bottom or vise-versa and make the decision based on the sequence representation up to now. This could be simply achieved using an LSTM layer. This approach delivered the best results. 

\textbf{Sequence Aggregation.} 
Given a sequence of probabilities, we can consider multiple aggregation strategies. A natural way to aggregate a sequence into a single result is using the $max$ value, e.g. a sequence is positive if a single fracture is found. The $max$ value could be unreliable due to noise; thus, smoothing is applied as a preceding step. An alternative approach is by considering multiple items in the sequence, either in a certain order or not. One possible ordering is simply the sequence of patches along the spine from top to bottom or vice versa. After experimenting different approaches as described in section \ref{sec:results}, we finally opted for the $max$ function.

\section{Experiments}

\subsection{Dataset}  
The data includes retrospectively collected CT chest and abdomen images acquired on different types of scanners  (Siemens, Philips and General Electric), with a vast representation of institutions, containing the vertebrae between T1 to L5. Patients were separated to independent training, tuning and test partitions, with 80\%, 10\%, and 10\% of the patients, respectively. Ground truth was determined by consensus agreement of three US Board Certified radiologists. Each Radiologist was provided with full-volume sagittal CT series and identified the presence or absence of vertebral compression fracture according to the Genant scale \cite{genant_1993}.
\par The training set consisted of 1,832 standard CT series of patients 50 years old and higher , with 613 (33\%) positive series and 1,219 (67\%)  negative series for VCF. Cardiac focused chest CT, lumbar spine, chest CT with lung windowing and PET-CT were excluded, since there is little value in opportunistic screening of vertebral compression fractures for this type of data.  
\par The  tuning  set,  used  for setting the  algorithm hyper-parameters consisted  of  311  series,  133 (42.8\%)  of  them  were  positive  for  VCF  and  178  (57.2\%)  were  negative. The tuning set distributed similarly to the training set and had the same exclusion criteria. Finally, the held-out validation set to test the algorithm performance was curated from  the  test  patients  partition,  and  consisted  of  346  series,  153  (44.2\%)  of them considered positive. This set distributed similarly to training and tuning sets.

\subsection{Implementation Details}
\textbf{Spinal Cord Localization} The model's validation set comprised of 200 additional axial slices unseen during training. The results were then validated by an expert. The expert reviewed the model's bounding box localization of the spinal cord and agreed with 98.99\% of the cases. 

\textbf{Image Augmentation.} Each spine patch ROI in the series sequence was preprocessed with the following additional augmentations: random horizontal and vertical flip 50\% of all the images and random rotation (up to $\pm$ 20 degrees).
Augmentations were chosen to align with commonly-used approaches (\cite{bar_2017}, \cite{peris_2019}) and limited to 50\% of all cases to ensure the usual appearance of the vertebrae is preserved.

\textbf{Mini-Batch Training.} For training, we use a learning rate of 1e-5 and batch size of 16. The weights were randomly initiated and trained for 2000 epochs with 150 iterations per epoch and an equal proportion of the positive and negative classes per batch. Models were trained using 2 NVIDIA Tesla K80 GPUs with 12 GB memory, using Keras and Tensorflow backend. We used the Adam optimizer, saving the model with the best ROC AUC on our validation set.

\subsection{Model Ensembles and Test Time Augmentation}
Our final model is made of an ensemble of three models and one Test Time Augmentation (TTA) per model (flip left/right or identity). In the ensemble the models output is averaged to a single score. To find the best ensemble an exhaustive search was performed on the hyper-parameters of the model including: patch input size, HU windowing, weight initialization approaches, learning rate and batch content. In model architecture we evaluated the following: (i) a sequence model with a LSTM classifier,(ii) a sequence model with max probability, and (iii) a shallow 3D convolution model. The best ensemble was found among the candidates in the pool with respect to each model ROC AUC on the tuning set. 

\section{Results}
\label{sec:results}

\begin{figure}
\centering
  \begin{subfigure}{.4\textwidth}
    \vspace{-0.1cm}
    \centering
    \includegraphics[width=.9\linewidth]{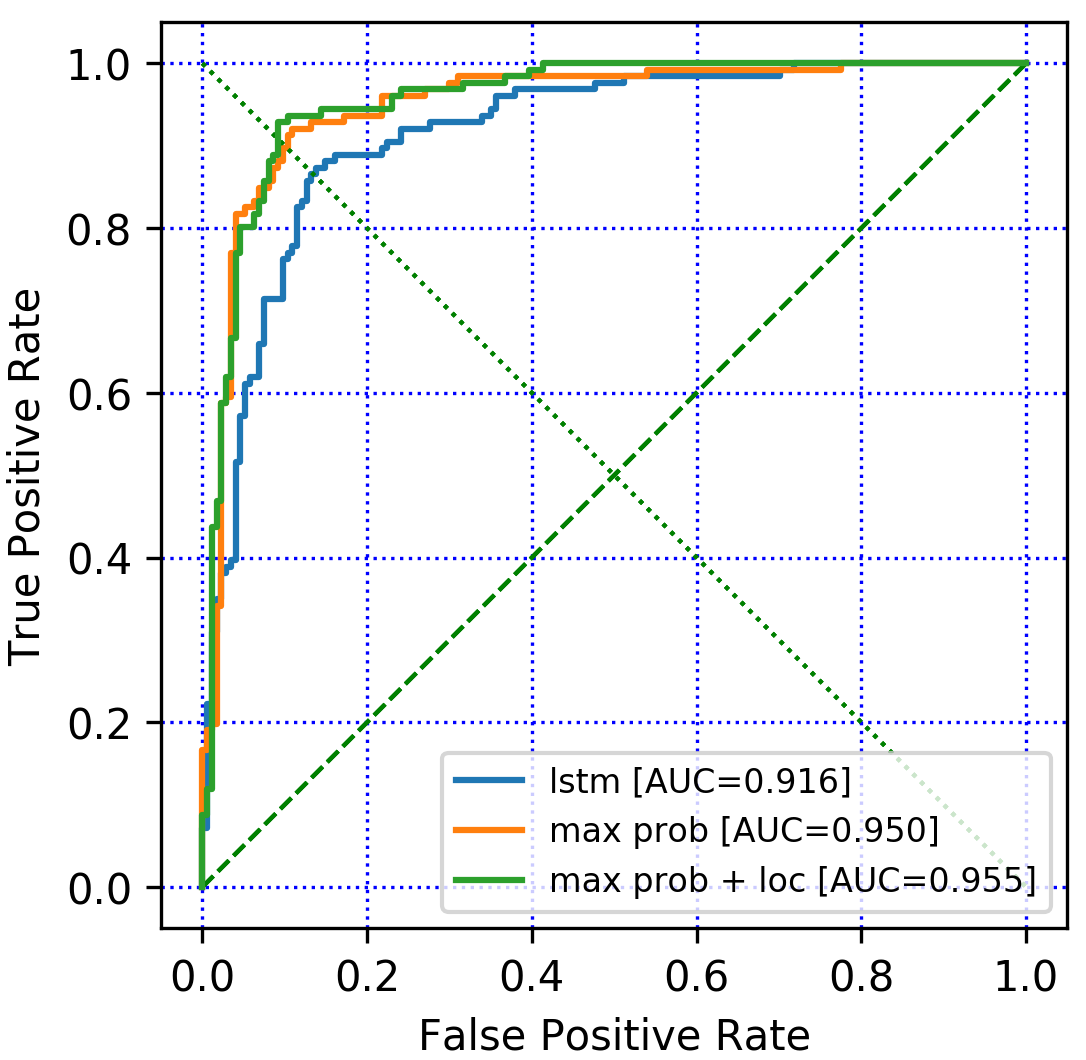}
    \captionsetup{margin=0.5cm}
    \caption{\label{fig:ROC_curves_experiments}Model experiments results. ROCs computed for different aggregation approaches on the tuning set.}
  \end{subfigure}%
  \begin{subfigure}{.4\textwidth}
    \centering
    \includegraphics[width=.9\linewidth]{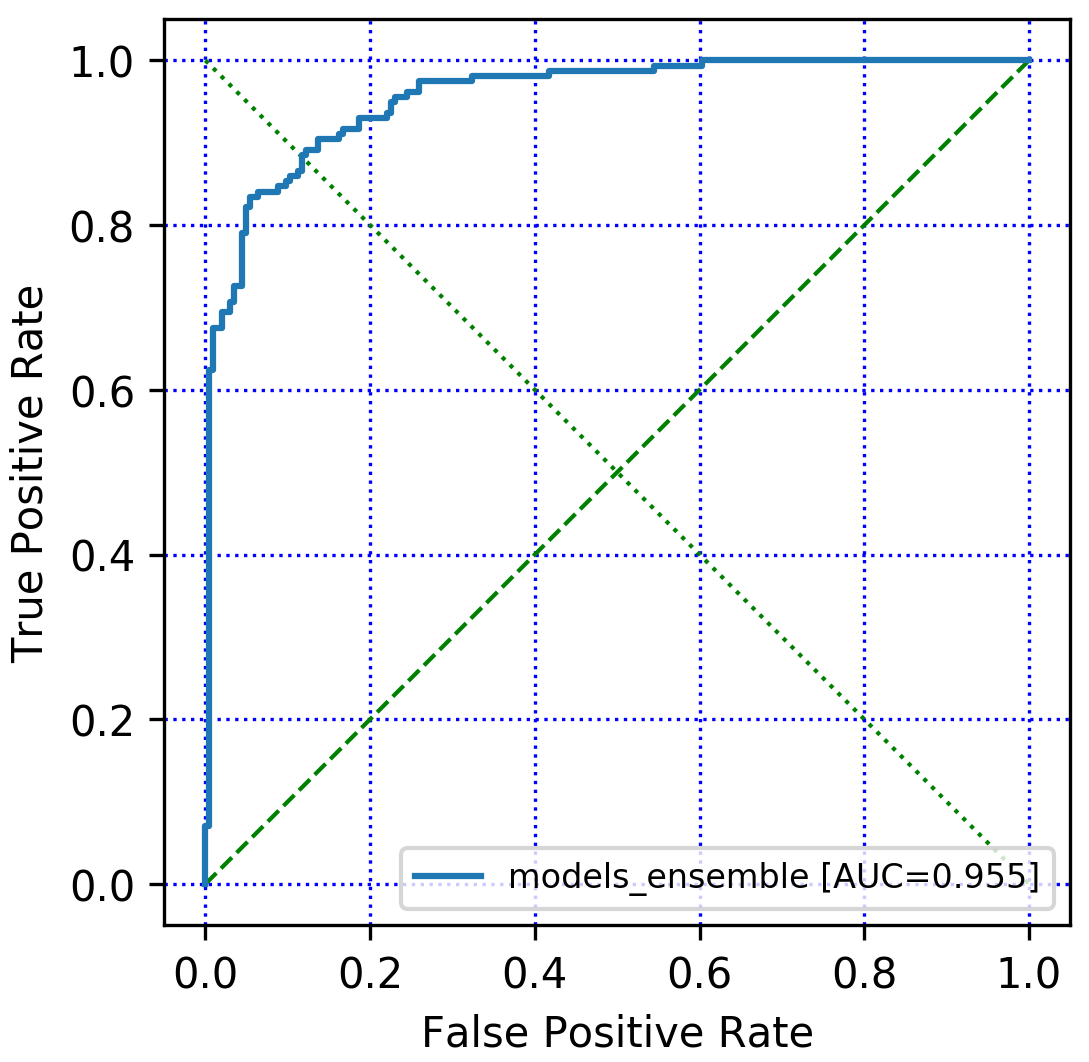}
    \captionsetup{margin=0.5cm}
    \caption{\label{fig:ROC_validation}Results on the held-out validation set. It plots the ROC curve of our final ensemble model.}
  \end{subfigure}%
\caption{\label{fig:sample_results}Experiments and test results.}
\end{figure}

We compared different strategies for series level classification working in the feature spaces of the 3D patches, using three classification approaches: (i) the maximum patch probability, (ii) the maximum patch probability with the patch dependant localization information, and (iii) a BiLSTM aggregation. Best results were obtained by maximizing the scores of the patch features, as shown in figure \ref{fig:ROC_curves_experiments}. This classification method achieved our best performance: 0.961 AUC ROC on the tuning set.

\par Figure \ref{fig:ROC_validation} shows our method patient-level fracture identification ROC
curve obtained on the held-out validation set. We also compared our proposed method to previously described methods in the literature, as shown in table \ref{tab:comp_lit}. We note that all these results have been reported using different validation sets. Based on this table, we can see that our proposed method is on par with the best AUC ROC reported so far \cite{nicolaes_2020}, but is trained and validated on a significantly larger and more diverse data set. 
\par
\begin{figure}[ht!]
\centering
\begin{subfigure}{.4\textwidth}
\includegraphics[scale=0.27]{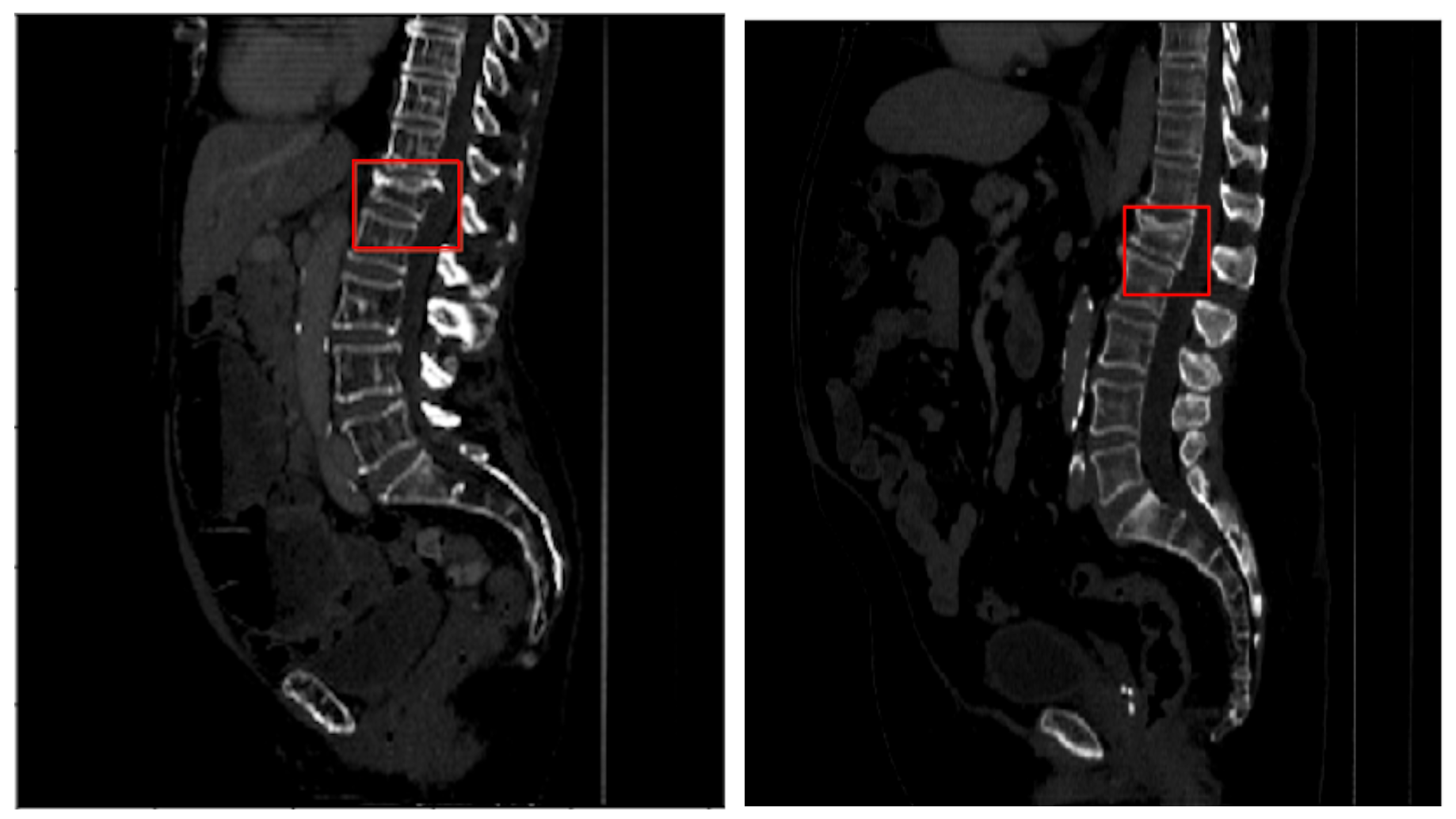}
\caption{\label{fig:a_tp} Correctly detected VCFs.}
\end{subfigure}
\begin{subfigure}{.4\textwidth}
\includegraphics[scale=1.1]{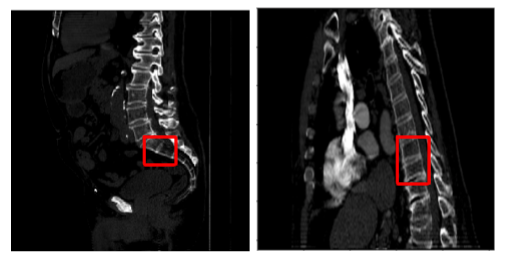}
\caption{\label{fig:b_fp}Incorrectly detected VCFs.}
\end{subfigure}
\caption{Illustration of the results. The VCF detected by the algorithm is displayed on the reconstructed mid-sagittal slice of the pre-processed 3D input image. The first two images in \ref{fig:a_tp} are the result of correctly detected VCFs, whereas the following two in \ref{fig:b_fp} show incorrectly detected ones. There, our system confused L5 and Schmorl nodes with VCFs.}

\label{fig:results}
\end{figure}

 Figure \ref{fig:results} provides illustration of both correctly and incorrectly detected VCFs, marked by a bounding box. During development, we observed that a common sources of false positive results were Genant Mild fractures, which represent a small range of vertebral height loss between 20-25\% and also obtain considerably lower agreement among clinicians \cite{genant_1993}. But, also due to confusion with other vertebral pathologies, such as Schmorl nodes.
\par Regarding performance, we report an average processing time per case of 61.36 s for the entire pipeline (standard deviation 43.16 s). These measurements were made on our test set using a CPU with 16 cores.

\begin{table}[!t]
  \setlength{\tabcolsep}{1.8mm}
  \centering{
  \begin{tabular}{|c|ccccc|}
  \hline
  & AUC & Sen & Spec & Data (\#cases) & Model type \\
    \hline\hline               
     Bar et al \cite{bar_2017} & $-$ & 0.839 & 0.938 & 3'701 & 2D \\
     Tomita et al.  \cite{tomita_2018} & 0.92 & 0.88 & 0.71 & 1'432 & 2D \\
     Valentinitsch et al. \cite{valentinitsch_2019} & 0.88  & 0.80 & 0.74 & 154 & 3D \\
     Nicolaes et al. \cite{nicolaes_2020} & 0.95 &  0.905 & 0.938 & 90 & 3D\\
    \hline
     Ours &  \textbf{0.955} & 0.822 & 0.951 & 2'489 & 3D \\
     \hline
  \end{tabular}}
\vspace{5pt}
  \caption{Comparison of our method to previously proposed approaches in the literature. Note “$-$” indicates data is not available.} 
  \label{tab:comp_lit}
 \end{table}

\section{Conclusions}
This study describes the development and testing of a novel end-to-end patient-level vertebral fracture identification system on CT. The system introduces a compact and effective 3D representation for VCF identification, a novel ROI extraction pipeline and leveraging a sequence to sequence architecture, obtaining SOTA results in this problem domain on a large data set. With to this architecture, location context and information from adjacent vertebrae are fused to the model. 
We expect the proposed system to improve VCF diagnosis in a clinical settings by pre-screening routine CT examinations and flagging suspicious cases. Future work, will focus on extending this approach to other modalities and osteoporosis applications. 

\clearpage
\bibliographystyle{ieeetr}
\bibliography{paper}

\end{document}